\documentclass[reprint,showpacs,superscriptaddress,floatfix,pre]{revtex4-1}
\usepackage{amsmath}
\usepackage{amssymb}
\usepackage{amsbsy}
\usepackage[dvips]{graphicx}

\def\eq#1{{Eq. (\ref{#1})}}

\begin{document}

\title{Hopping charge transport in organic materials}

\author{S.V. Novikov}
\email{novikov@elchem.ac.ru}
\affiliation{A.N. Frumkin Institute of
Physical Chemistry and Electrochemistry, Leninsky prosp. 31,
119991 Moscow, Russia}

\begin{abstract}
 General properties of the transport of charge carriers (electrons and holes) in disordered organic materials are discussed. It was demonstrated that the dominant part of the total energetic disorder in organic material is usually provided by the electrostatic disorder, generated by randomly located and oriented dipoles and quadrupoles. For this reason this disorder is strongly spatially correlated. Spatial correlation directly governs the field dependence of the carrier drift mobility. Shape of the current transients, which is of primary importance for a correct determination of the carrier mobility, is considered. A notable feature of the electrostatic disorder is its modification in the vicinity of the electrode, and this modification takes place without modification of the structure of the material. It is shown how this phenomenon affects characteristics of the charge injection. We consider also effect of inter-charge interaction on charge transport.
\end{abstract}

\maketitle

\section{Introduction}
Charge carrier transport in organic materials is a vibrant area of research for over three decades. To a  large degree this attention is motivated by the constantly growing application of organic materials in various electronic devices. In this paper we will completely put aside this aspect of the organic electronics and focus entirely on the fundamental properties of the conductivity of organic materials. Reviews of advances in the application of electronic organic devices may be found in recent books \cite{Klauk:book,Brabec:book,Bernards:book}. In addition, we do not consider properties of highly conductive organic polymers (doped polyaniline, polyacetylene and others) and limit our attention to more traditional materials having low conductivity. By the usual classification they may be considered as amorphous semiconductors with wide bandgap. These materials are polymeric or low molecular weight organic glasses. Typical example of the conducting organic glass is a material created by doping of the inert polymer binder (polycarbonate or polystyrene) with molecules, providing conducting properties (mostly, aromatic amines, hydrozones, or nitriles); mass fraction of the dopant in most cases falls in the range  30\%--100\% \cite{Borsenberger:book}.

Most important features of the charge transport directly follow from the basic structural features of organic glasses. These glasses are molecular materials with rather weak interactions between molecules and, at the same time, they have significant disorder in positions and orientations of molecules. This means that all relevant states are localized and charge carrier transport occurs by the hopping mechanism \cite{Shklovskii:book}. It is well known that for the hopping transport the disorder in the material is of paramount importance. Traditionally, total disorder in amorphous materials is subdivided into energetic and positional disorders. The energetic disorder includes random fluctuations of the positions of energetic levels of molecules, while positional disorder describes fluctuations of the positions of molecules that affect hopping rates between transport sites but do not lead to the fluctuations of energies. There is a general  agreement in the community that the energetic disorder is of primary importance for the hopping transport (at least for the case where the concentration of transport sites is not too low) \cite{Bassler:15}. In this paper we consider general statistical properties of the energetic disorder in organic glasses, how they affect various features of the hopping transport in organic materials,  and provide comparison with well-established experimental facts.

\section{Basics of hopping transport in amorphous materials: density of states}

Typically, a hopping model is constructed as a regular cubic lattice with the lattice scale $a$, its sites are randomly occupied by transport molecules with the occupation fraction $c$, and a random energy $U_i$ is assigned to each site. In most cases the hopping rate has a Miller-Abraham (MA) form \cite{Miller:745}
\begin{equation}\label{MA_rate}
p_{i\rightarrow j}=\nu_0 \exp(-2\gamma r_{ij})\begin{cases}
    \exp\left(-\frac{U_j-U_i}{kT}\right), &U_j-U_i > 0\\1, &U_j-U_i < 0
\end{cases}
\end{equation}
where $\nu_0$ is some characteristic frequency of hops, $r_{ij}=|\vec{r}_j-\vec{r}_i|$, and $\gamma$ is a wave function decay parameter for transport sites; in organic materials $\gamma a\simeq 5-10$ \cite{Bassler:15,Pasveer:206601}. If an electric field $E$ is applied, then the random energy includes an additional term $-e\vec{r}_i\vec{E}$. Distribution of random energies is typically considered  having a Gaussian \cite{Bassler:15,Pope:1328} or  exponential form \cite{Arkhipov:7909}. In this paper we limit our consideration to the Gaussian density of states (DOS). This particular form is widely recognized as more suitable for description of organic amorphous solids \cite{Bassler:15}. Apart from the experimental evidence, the Gaussian DOS naturally arises in simple but reasonable models of disordered organic materials. For example, the simplest model of the polar glass is the lattice with sites occupied by randomly oriented dipoles \cite{Dieckmann:8136,Novikov:877e}. If fraction of occupied sites is not too low ($c\simeq 1$), then the resulting DOS $P(U)$ has a Gaussian form with the width
\begin{equation}\label{sigma}
    \sigma=2.35\frac{epc^{1/2}}{\varepsilon a^2},
\end{equation}
 here  $p$ is the dipole moment, and $\varepsilon$ is the dielectric constant of the medium \cite{Novikov:877e} (numeric coefficient $2.35$ is specific for the simple cubic lattice). For typical values of $p$, $\varepsilon$, and $c$ the magnitude of disorder $\sigma \approx 0.05-0.1$ eV. For $c\ll 1$ there is an  intermediate asymptotics
\begin{equation}\label{low_c}
    P(U)\propto \frac{1}{U^{5/2}}, \hskip10pt \frac{epc^{2/3}}{\varepsilon a^2}\ll U \ll \frac{ep}{\varepsilon a^2}
\end{equation}
but in practice this regime cannot be observed due to inevitable additional contribution to the DOS from other sources of disorder \cite{Novikov:877e}.

Particular shape of DOS directly governs the mobility temperature dependence $\mu(T)$ (for a low field region). Gaussian DOS leads to
\begin{equation}\label{mu_T}
    \ln \mu \approx -A\left(\frac{\sigma}{kT}\right)^2,
\end{equation}
where $A\simeq 1$ is some constant, and this particular dependence is believed to be the most properly suited for description of experimental data  \cite{Schein:4287,Borsenberger:9,Schein:773}. Roots of  \eq{mu_T} may be understood if we calculate the position of the maximum of the density of \textit{occupied} states
\begin{eqnarray}\label{max}
    P_{\textrm{occ}}(U)\propto P(& U)\exp(-U/kT)\propto
    \exp\left(-\frac{U^2}{2\sigma^2}-\frac{U}{kT}\right), \nonumber    \\ & U_{\textrm{max}}=-\frac{\sigma^2}{kT},
\end{eqnarray}
in the Gaussian DOS, and then use $U_{\textrm{max}}$ as an equivalent of the temperature-dependent activation energy.

Motivated by this evidence, B{\"a}ssler suggested a Gaussian Disorder Model (GDM) for the description of hopping charge transport in amorphous organic materials \cite{Bassler:15}. This model was claimed to be a universal model suitable to describe charge transport in any disordered organic material. Main ingredients of the GDM are the assumption of the Gaussian DOS and validity of the Miller-Abrahams hopping rate (\ref{MA_rate}). It was assumed also that the distribution of random energies has no spatial correlation at all; the correlation function is $\left<U(\vec{r}_i)U(\vec{r}_j)\right>=\sigma^2\delta_{ij}$.
The major part of results has been obtained for the GDM by means of Monte Carlo simulation. For example, it was found that for the GDM $A\approx 4/9$ in \eq{mu_T}.

\begin{figure}[tbh]
\begin{center}
\includegraphics[width=3.4in]{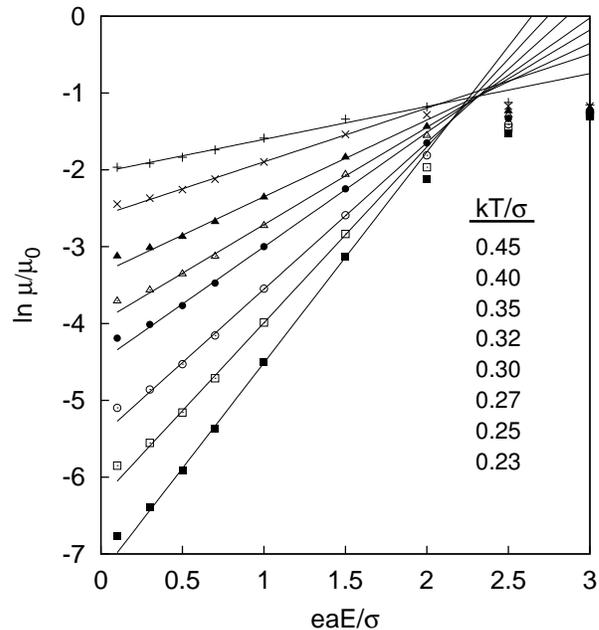}
\end{center}
\caption{Field dependent mobility in the GDM for different values of $kT/\sigma$ (from the top curve downward); $\mu_0=ea^2\nu_0/\sigma\exp\left(-2\gamma a\right)$. Straight lines show best fits for the linear regions of the curves.}
\label{GDM_E}
\end{figure}

Another important parameter affecting mobility is an applied electric field $E$. For a long time it was known that organic glasses do demonstrate a strong field dependence of the mobility \cite{Gill:5033,Borsenberger:9}, and the so called Poole-Frenkel (PF) dependence
\begin{equation}\label{mu_E}
\ln\mu\propto\textrm{const}+\sqrt{E}
\end{equation}
is usually a good description of the experimental data. While the GDM fairy well describes the mobility temperature dependence, it totally fails in describing its field dependence. There are claims that the GDM \textit{do} reproduce the PF dependence \eq{mu_E}, though in a limited field range \cite{Bassler:15}. This statement is not true. Field dependence of $\mu$ in GDM for low and moderate fields is much better described by a linear dependence
\begin{equation}
\ln\frac{\mu}{\mu _{0}}= -0.38\left( \frac{\sigma }{kT} \right)
^{2} +1.17 \left( \frac{\sigma}{kT}-2.05
 \right) \frac{eaE}{\sigma}.  \label{mu_GDM}
\end{equation}
This equation summarize the results of the Monte Carlo simulation (see Fig. \ref{GDM_E}), more thoroughly described elsewhere \cite{Novikov:3047}. In fact, the parabolicity of the plot $\ln\mu$ vs $E^{1/2}$ is clearly seen even in Fig. 7 of the B\"{a}ssler's paper \cite{Bassler:15}. Close inspection of Fig. \ref{GDM_E} (as well as Fig. 7 of paper \cite{Bassler:15}) shows that the only field range that can mimic the PF dependence (\ref{mu_E}) is the region $eaE/\sigma \gtrsim 1$,  where the mobility curve, plotted as  $\ln\mu$ -- $E$, begins to deviate from the straight line. The reason for this deviation is the use of the particular hopping rate, that is the MA rate: it has a property that for very strong electric field, where almost all hops occur downward in energy, carrier velocity is saturated and, hence, the mobility begins to decay with $E$ as $1/E$. By its very nature, this mechanism cannot provide a good linearity of the dependence $\ln\mu$ vs $E^{1/2}$ for the range $10^5-10^6$ V/cm, where it is routinely observed \cite{Borsenberger:9} (if $\sigma=0.1$ eV, then for the typical scale $a=1$ nm $eaE/\sigma\approx 1$ for $E=1\times 10^6$ V/cm). In some materials the PF dependence was tested even for much wider range, such as $8\times 10^3 - 2\times 10^6$ V/cm \cite{Schein:686}.

Inability to describe the mobility field dependence indicates that some important element is missing in the GDM. We are going to find this missing element.

\begin{figure}[tbh]
\begin{center}
\includegraphics[width=3.4in]{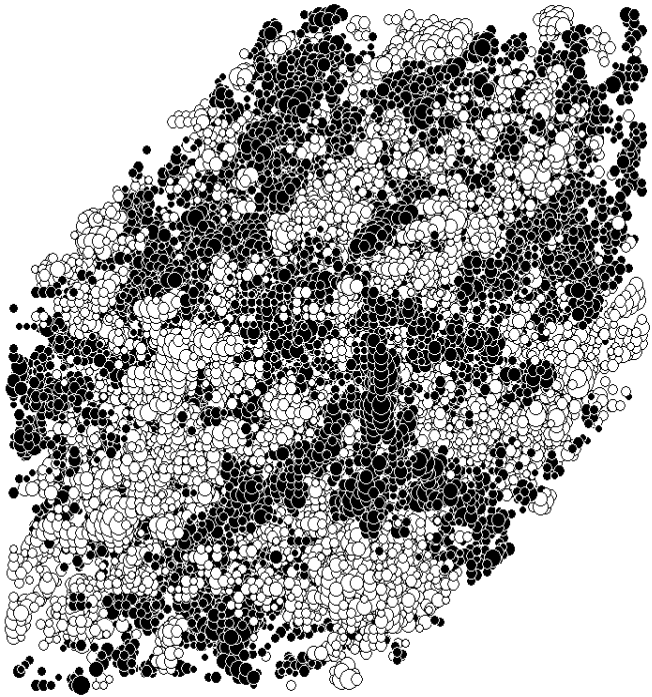}
\end{center}
\caption{Distribution of site energies $U$ in the lattice model of dipolar glass. A sample with the  size of $50\times 50\times 50$ lattice sites is shown. Black and white spheres represent the sites with positive and negative values of $U$, correspondingly, while the radius of a sphere is proportional to the absolute value of $U$. Sites with small absolute values of $|U|$ (less than $\sigma$) are not shown for the sake of clarity. }
\label{clusters}
\end{figure}

\section{Organic glasses: kingdom of spatial correlations}
\subsection{Long range spatial correlations of the energy landscape in organic glasses}

The GDM is based on the assumption of the non-correlated nature of the random energy landscape in organic materials. It turned out that this assumption proved to be spectacularly wrong in organic glasses. Indeed, we already mentioned that organic glasses have very low concentration of free intrinsic carriers and, hence, do not provide effective screening of electrostatic interactions. At the same time, they have high concentration of permanent dipoles and quadrupoles. These molecules provide long range (unscreened) random contributions to the total random energy of charge carriers. Sum of random terms, slowly decaying in space, inevitably produce strongly correlated random energy landscape. For dipoles the corresponding correlation function decays as  \cite{Novikov:14573}
\begin{equation}\label{Cd}
C_{d}(\vec{r})\approx A_d\sigma^2 \frac{a}{r},
\end{equation}
 while for quadrupoles \cite{Novikov:181}
\begin{equation}\label{Cq}
C_{q}(\vec{r})\approx A_q\sigma^2 \left(\frac{a}{r}\right)^3.
\end{equation}
Dimensionless parameters $A_d$ and $A_q$ equal to $0.76$ and $0.5$, correspondingly, for the simple cubic lattice \cite{Dunlap:80,Novikov:181}. Quite probably, the correlation functions of dipolar or quadrupolar type are the most common ones in organic glasses, because the model of randomly oriented and located dipoles provides  a good approximation for the polar organic glasses, and the corresponding model of quadrupoles gives a reasonable model for non-polar organic glasses.

Long range correlations mean that the random energy landscape in organic glasses has a natural cluster structure: sites with close values of $U$ tend to group together (see Fig. \ref{clusters}). Abundance of  clusters in organic glasses may be characterized by the cluster distribution on size: for the dipolar glass the asymptotics for the number of clusters $n_s$ having $s$ sites decays as
\begin{equation}\label{clust_DG}
    \ln n_s\propto -s^{1/3},
\end{equation}
while for the non-correlated GDM
\begin{equation}\label{clust_GDM}
    \ln n_s\propto -s,
\end{equation}
which gives a huge difference for $s\gg 1$ \cite{Novikov:41139,Novikov:241}.

\subsection{How correlations affect charge mobility dependence}
\label{sect_transport}

Correlation nature of the energy landscape $U(\vec{r})$ directly dictates major features of the charge transport and injection in organic glasses. For example, the field dependence of the quasi-equilibrium mobility can be understood from the following simple consideration (for more thorough consideration see Ref. \cite{Dunlap:437}). Suppose that the carrier is located at the bottom of the potential well with the energy $U(0)$. Mobility is determined by the typical time for the carrier to reach a saddle point with the energy $U(\vec{r})-e\vec{E}\vec{r}$, where $r$ is the distance to the saddle point from the bottom of the well. That time can be estimated as
\begin{equation}
\label{typical_time}
t\simeq t_0 \exp\left[\frac{U(\vec{r})-U(0)-e\vec{E}\vec{r}}{kT}\right],
\end{equation}
where $t_0\simeq 1/\nu_0$, and the average time for the Gaussian random landscape $U(\vec{r})$ is
\begin{eqnarray}
\label{typical_time_avg}
\left<t\right>\simeq t_0 \left<\exp\left[\frac{U(\vec{r})-U(0)}{kT}\right]\right>\exp\left(-\frac{e\vec{E}\vec{r}}{kT}\right)= \\ \nonumber
=t_0\exp\left\{\frac{\left<\left[U(\vec{r})-U(0)\right]^2\right>}{2(kT)^2}-\frac{e\vec{E}\vec{r}}{kT}\right\}= \\ =t_0 \exp\left[\frac{C(0)-C(\vec{r})}{(kT)^2}-\frac{e\vec{E}\vec{r}}{kT}\right].\nonumber
\end{eqnarray}
Assuming charge transport in the dipolar glass (DG) with the correlation function (\ref{Cd}), we can calculate the critical size of the potential well that provides the maximal escape time
\begin{equation}
\label{critical}
\frac{d\left<t\right>}{dr}=0, \hskip10pt r_{\rm cr}=\sigma
\left(\frac{aA_d}{eEkT}\right)^{1/2},
\end{equation}
and the mobility is estimated as
\begin{equation}
\label{mob_crit}
\mu\propto 1/\left<t\left(r_{\rm cr}\right)\right>\propto \exp\left[-\left(\frac{\sigma}{kT}\right)^2+\frac{\sigma}{kT}\left(\frac{eaA_dE}{kT}\right)^{1/2}\right].
\end{equation}
This result provides a leading asymptotics of the exact solution of 1D transport problem \cite{Dunlap:542,Parris:5295}.
Typical magnitude of the slope of the mobility field dependence $\ln\mu$ vs $E^{1/2}$, calculated using \eq{mob_crit}, agrees well with the estimations that follow from the experimental data for $\sigma$, obtained from the low field mobility temperature dependence \cite{Borsenberger:9}. This means that in polar materials the dipolar $\sigma$ (estimated from the mobility field dependence) provides a dominant part of the total $\sigma$ (estimated from the mobility temperature dependence). Extensive computer simulation of the 3D transport generally confirms \eq{mob_crit} and only modifies numeric parameters in this relation. Results of the simulation may be summarizes as a phenomenological relation
\begin{equation}
\ln\frac{\mu}{\mu _{0}}= -\left( \frac{3\sigma }{5kT} \right)^{2} +C_{0} \left[ \left( \frac{\sigma }{kT}\right)^{3/2}
-\Gamma\right]\sqrt{eaE/\sigma },
\label{eq5}
\end{equation}
where $C_0 \approx 0.78$, and $\Gamma \approx 2$ \cite{Novikov:4472}.

In a more general case of the algebraic correlation function
\begin{equation}\label{Cn}
    C(\vec{r})=A_n\sigma^2\left(\frac{a}{r}\right)^n
\end{equation}
the mobility in 1D case and strong disorder $\sigma/kT\gg 1$ is \cite{Novikov:2532}
\begin{equation}
\ln\frac{\mu}{\mu_0} \approx-\left(\frac{\sigma}{kT}\right)^2+\left(1+\frac{1}{n}\right)
\frac{\sigma}{kT}\left(\frac{A_n n\sigma}{kT}\right)^{\frac{1}{n+1}}\left(\frac{eaE}{\sigma}\right)^{\frac{n}{n+1}}.
\label{1D}
\end{equation}
For the GDM formally $n\rightarrow\infty$ and the result agrees well with the leading field-dependent term in \eq{mu_GDM}.

\begin{figure}[htb]
\begin{center}
\includegraphics[width=2.9in]{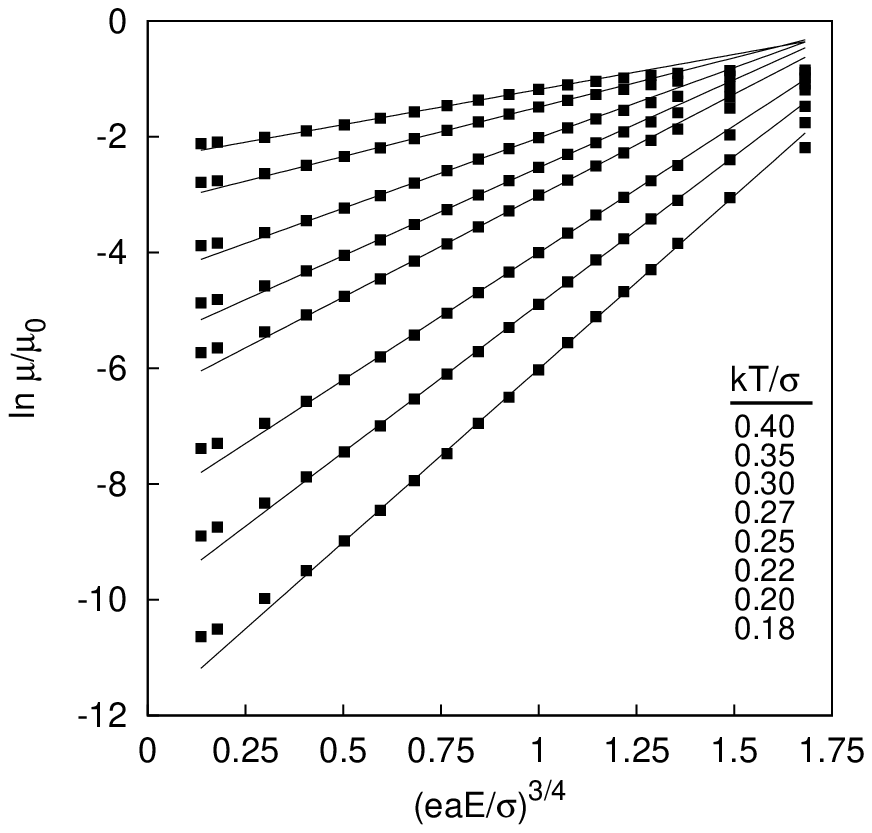}
\end{center}
\centering a
\begin{center}
\includegraphics[width=2.9in]{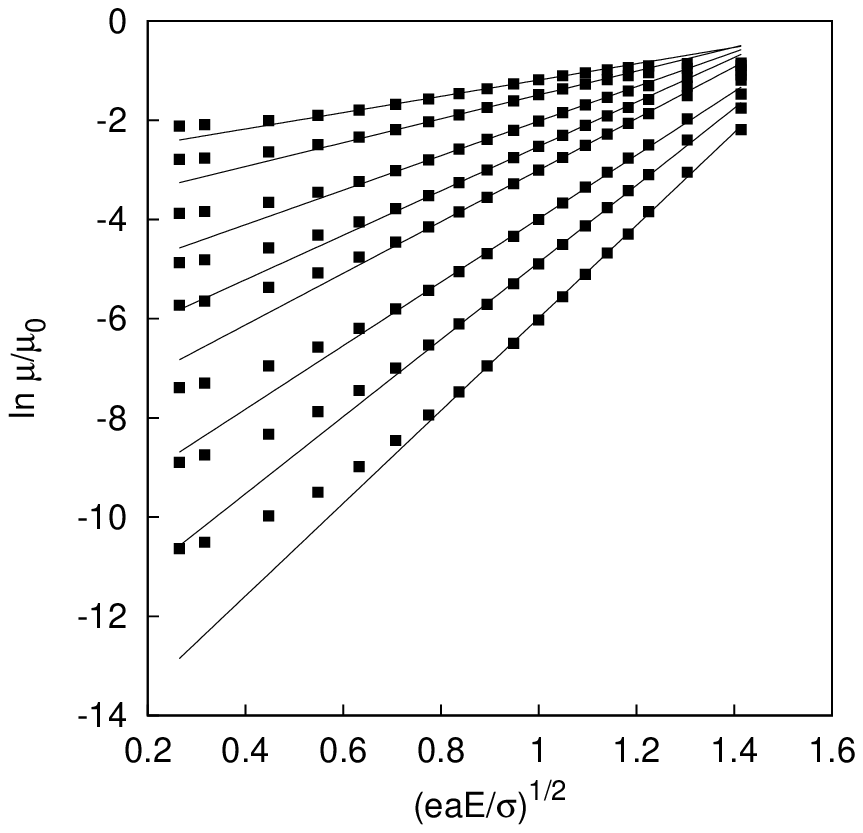}
\end{center}
\centering b
\caption{Mobility field dependence in the QG model for different values of $kT/\sigma$ (from the top curve downward); straight lines in (a) indicate the fit for \eq{QGmu}. Plot of the simulation data in the usual PF presentation $\ln\mu$ vs $E^{1/2}$ (b) demonstrates much stronger deviation from the linearity in the weak field region (straight lines serve as a guide for an eye).}
\label{QG_E}
\end{figure}

This analysis indicates not only that the GDM fails to capture a very important characteristic of organic glasses (spatial correlations), but also that there is no unified mobility field dependence for different classes of organic materials. The functional form of the mobility field dependence depend on the spatial decay of the correlation function and is, indeed, different in polar and non-polar organic glasses \cite{Novikov:89}. This conclusion of 1D analysis is in good agreement with the result of 3D simulation for the model of quadrupolar glass (QG), serving as a reasonable model of the non-polar organic glass (Fig. \ref{QG_E}). Simulation data can be reasonably well described by the equation
\begin{equation}
\ln\frac{\mu}{\mu_0}=-0.37\left(\frac{\sigma}{kT}\right)^2+C_Q\left[\left(\frac{\sigma}{kT}\right)^{5/4}-
\Gamma_Q\right]\left(\frac{eaE}{\sigma}\right)^{3/4},
\label{QGmu}
\end{equation}
with $C_Q\approx 0.87$ and $\Gamma_Q\approx 1.91$ \cite{Novikov:954}.

For the non-correlated Gaussian landscape mobility is controlled by the carrier release from the deep states to the neighbor sites having higher energy, hence the shift of the carrier energy in the applied field lead to the linear field dependence
\begin{equation}
\ln\mu\propto eaE/kT,  \label{mu_GDM_simple}
\end{equation}
which again agrees well with \eq{mu_GDM} and the limiting case of \eq{1D}.

\begin{figure}[tbph]
\begin{center}
\includegraphics[width=3.2in]{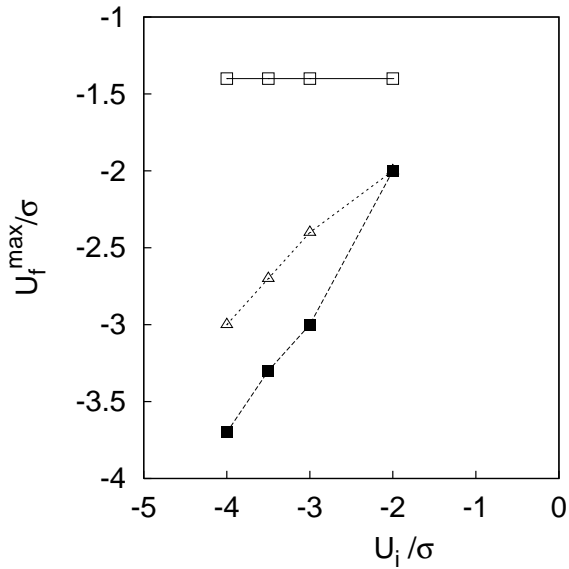}
\end{center}
\caption{Position of the maximum of  distribution of final carrier energy $U_f$ in DG ($\blacksquare$), QG ($\vartriangle$), and GDM ($\square$), correspondingly, in dependence on the initial carrier energy $U_i$ for $kT/\sigma = 0.25$. Lines are provided for the convenience.}
\label{maxfig}
\end{figure}

An interesting difference between correlated energy landscapes and the GDM is a non-existence of the so-called transport energy in organic glasses \cite{Novikov:387}. This conception was extensively used for an analysis of the hopping transport in amorphous materials \cite{Baranovskii:283,Baranovskii:2699,Arkhipov:7514}. Transport energy serves as an analogue of the mobility threshold for the hopping transport. It emerges as a result of the competition of two opposite tendencies for a carrier, escaping from low energy sites: fast decay of the hopping rate \eq{MA_rate} with distance facilitates hops to the nearest neighbors, while the the probability to find a neighbor with not too high energy (and, hence, not too low hopping probability) increases with distance. As a result, in the GDM there is an optimal hopping distance and the optimal final energy  (transport energy) for a carrier, which does not depend on the initial carrier energy. In the correlated landscape site, close in space, are the sites, close in energy, too, and the optimal final energy does not exist (the most probable final energy shifts with the variation of the initial carrier energy \cite{Novikov:387}, see Fig. \ref{maxfig}).

The most important conclusion of this section is that the functional form of the mobility field dependence is directly dictated by spatial correlations in the particular organic material. The hope to provide a single universal formula for $\mu(T,E)$ is futile, and the major watershed divides the polar and non-polar organic materials. At the same time, analysis of experimental data indicates that a reliable determination of the mobility field dependence is not a trivial task. For all non-polar organic glasses rather limited experimentally tested field range is typical (one order of magnitude or even more narrow \cite{Borsenberger:226,Borsenberger:555,Borsenberger:9,Sinicropi:331,Heun:245}).
Taking into account inevitable experimental errors, this means that the reliable discrimination between the PF dependence (\ref{mu_E}) and the true quadrupolar dependence (\ref{QGmu}) is hardly possible (a good example is discussed in the paper \cite{Novikov:2532}).

\section{Current transients: conception of universality}

A most important method for the experimental determination of the hopping mobility in organic materials is a time-of-flight (TOF) method \cite{Borsenberger:book}. Here the experimental sample is a slab of organic material with thickness $L$ sandwiched between two electrodes. Typically, electron-hole pairs  are generated in the vicinity of one electrode (though a variant with the uniform bulk generation is used, too \cite{Tyutnev:523}, but for this variant some serious difficulties in the interpretation of the experimental data take place \cite{Novikov:534}), carriers of one sign are instantly absorbed by the electrode, and carriers of the opposite sign drift through the slab to the collecting electrode under the action of the applied field $E$. In the TOF method the raw experimental data is the time dependence of the current $I(t)$, generated by drifting carriers (current transient).

In most cases transient demonstrates an initial short spike, indicating a spatial and energetic relaxation of carriers, then a plateau (at this stage carriers move with the almost constant velocity $v$), and finally a fast decay, indicating an arrival of carriers to the collecting electrode. These are typical features of so-called non-dispersive (or quasi-equilibrium) transport, and the most important property is the independence of a mean velocity $\left<v\right>$ on $L$, well established in experiments \cite{Stolka:4707}. Mobility is calculated as
\begin{equation}\label{mu_def}
\mu=\frac{\left<v\right>}{E}\approx\frac{L}{t_dE},
\end{equation}
where $t_d$ is some characteristic drift time. In most papers $t_d$ was chosen as a time $t_0$ of the intersection of the asymptotes to the plateau and tail of the current, while in some papers $t_d$ is the time $t_{1/2}$ of the current to decay to one half of the plateau value.

In some cases (mostly for low temperature) plateau of the transient is not well defined and current demonstrates monotonous decay. This is the case of the dispersive (non-equilibrium) transport and the shape of the transient is analyzed in double logarithmic coordinates $\ln I$ vs $\ln t$ according to the popular Scher-Montroll model \cite{Scher:2455}
\begin{equation}\label{dispersive}
I(t)\propto\begin{cases}
    t^{-(1-\alpha)}, \hskip10pt t < t_T,\\t^{-(1+\alpha)}, \hskip10pt t > t_T,
\end{cases}
\end{equation}
here $t_T\propto L^{1/\alpha}$ is some characteristic time and $0< \alpha < 1$ is the dispersive parameter. In this regime the mean velocity depends on $L$ as $\left<v\right>\propto L/t_T\propto L^{1-1/\alpha}$.

If we discuss charge transport in term of mobility, then all we need from the experiment is the drift time $t_d$ (be it $t_0$, $t_{1/2}$, or $t_{T}$). Naturally, the whole shape of the transient could provide additional valuable information about the transport mechanism. In fact, sometimes even a proper discussion of the mobility field dependence requires a clear distinction of the mobilities, obtained with $t_0$ and $t_{1/2}$, i.e. an explicit use of some information about shape of the transient \cite{Novikov:444}. This remark explains the significance of the parameter
\begin{equation}\label{W}
    W=\frac{t_{1/2}-t_0}{t_{1/2}},
\end{equation}
which provides a simplest robust integral characteristic of the shape. Other important features of the transients are short and (especially) long time asymptotics (the initial spike and tail of the transient).

All theoretical approaches indicate that for the Gaussian DOS a constant (independent of time) velocity $v=\left<v\right>$ does eventually emerge and the quasi-equilibrium steady state is achieved \cite{Bassler:15}; for this reason in the paper we limit our  consideration to quasi-equilibrium non-dispersive transients. Experimental studies show that a general picture could be summarized in the following way. For high temperature $\sigma/kT \lesssim 3$ parameter $W$ decreases with $L$ as $W\propto 1/L^{1/2}$ \cite{Yuh:539,Borsenberger:967}, which is a fingerprint of the classical diffusion. Indeed, it was found that   the usual classical diffusion is a good approximation for a description of the shape of transients  in this situation \cite{Hirao:1787,Hirao:4755}.

For lower temperatures current transients have a different form \cite{Schein:175,Borsenberger:967}: for such transients $W(L)\approx {\rm const}$. Hence, if the time scale is properly re-scaled, then transients for different $L$ approximately collapse to a single universal curve (for this reason this phenomenon was dubbed universality). It was found that universality with respect to $L$ usually means the universality with respect to $E$, but not $T$ \cite{Schein:175}.

Previously, the very conception of universality has been invariably attributed with the dispersive transport. Schein et al. \cite{Schein:175} specifically emphasized that their's data suggest a universal behavior for the non-dispersive transport with the mean carrier velocity, which does not depend on $L$. This fact clearly indicates that a common perception (see, e.g. \cite{Hirao:1787}) that a transport with the well-defined constant velocity is invariably the transport, described by the usual diffusion equation, is certainly not true for organic glasses. In fact, this phenomenon has been well established long ago for some models (see the excellent review \cite{Bouchaud:127}), but did not attract much attention in the transport community.

Unfortunately, we have no reliable analytic results for the shape of transients in the random medium in 3D case. Most interesting exact results have been obtained for 1D transport \cite{Parris:2803}. This paper was specifically devoted to the study of the diffusing coefficient $D$ in correlated Gaussian energy landscape with DG correlation function \eq{Cd}. It was found that the Einstein relation
\begin{equation}\label{ER}
D =\frac{\mu kT}{e}
\end{equation}
does not hold, but the modified Einstein relation
\begin{equation}\label{mER}
D =\frac{kT}{e}\frac{\partial \left<v\right>}{\partial E}
\end{equation}
is valid, which transforms to \eq{ER} for $\mu(E)={\rm const}$. Yet the major result of the paper \cite{Parris:2803} is the very existence of the diffusion coefficient for all $T$ and $E$ in 1D transport. This is a drastic contradiction with the universality, found in experimental papers \cite{Schein:175,Borsenberger:967}.

\begin{figure}[tbph]
\begin{center}
\includegraphics[width=3.4in]{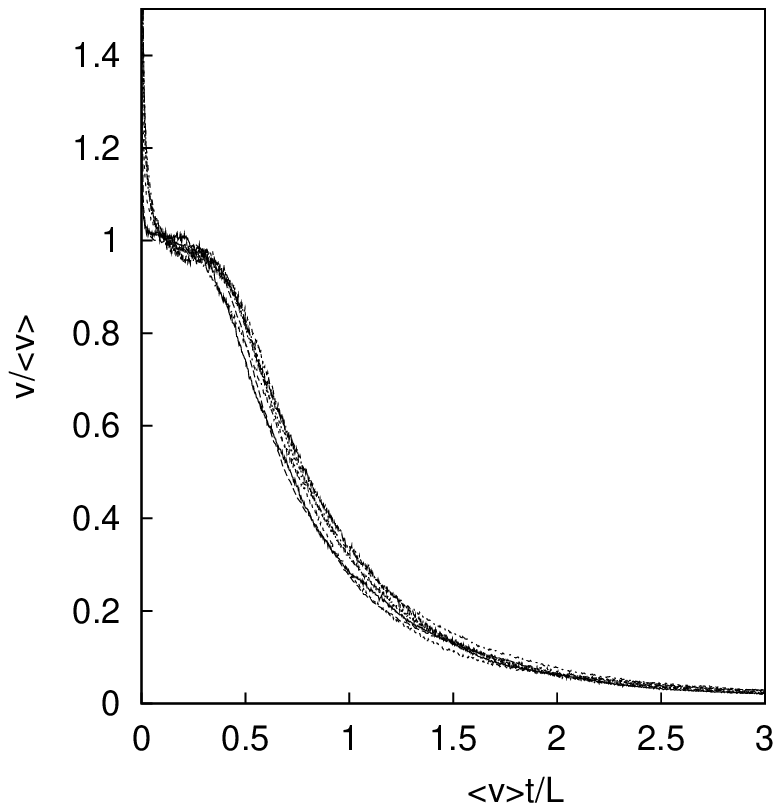}
\end{center}
\centering a
\begin{center}
\includegraphics[width=3.4in]{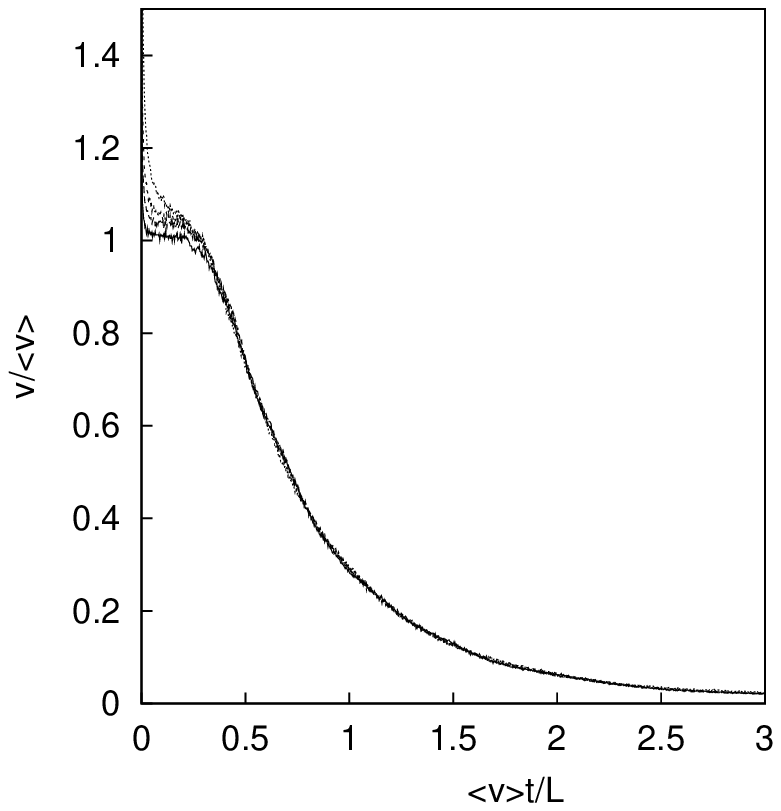}
\end{center}
\centering b
\caption{Test of the universality of current transients for the DG model; $kT/\sigma=0.19$ and $c=0.3$. Carrier velocity $v$ is directly proportional to $I(t)$. a) Transients for the electric field in the range 0.02 -- 1.2 $eaE/\sigma$ and $L=20,000a$. b) Transients for $eaE/\sigma=0.02$ and $L$ varying from 2,000 to 20,000 lattice scales.} \label{universal}
\end{figure}

Monte Carlo simulation agrees well with the experiments. Indeed, for high temperature the transients are diffusive \cite{Novikov:391}, while for the low temperature a universal behavior emerges (Fig. \ref{universal}). Note, that in our case the transport is definitely non-dispersive because a single curve is produced by re-scaling the time as $t\rightarrow t\left<v\right>/L$ and velocity as $v \rightarrow v/\left<v\right>$ (the later one is needed only for checking of the $E$-universality). Is the universality a distinct feature of the 3D transport or  true diffusive behavior still emerges in 3D case for a very long time, remains an open question. It is worth to note that according to the simulation data, the modified Einstein relation \eq{mER} does not hold for the 3D transport in the DG model even for high temperature \cite{Novikov:391}.

\section{Energetic disorder near the electrode}

We already noted that the total energetic disorder in organic glasses is mainly the electrostatic disorder, provide by randomly located dipoles and quadrupoles. For this reason, it demonstrates a very unusual property: modification of the statistical properties of the disorder near the conducting electrode without any modification of the structure of the material.


In fact, in any case structure of the organic material in the vicinity of the electrode should be quite different from the bulk structure. We should expect different packing of spacious organic molecules, accumulation of impurities, partial degradation of the organic material etc. For all these reasons the energetic disorder at the electrode is very different from the bulk disorder (typically, it is greater than the bulk disorder).

Yet in organic glasses there is an opposite general contribution, leading to the decrease of the electrostatic disorder at the electrode. The electrostatic energetic disorder is directly proportional $U(\vec{r})=e\varphi(\vec{r})$ to the disorder in the spatial distribution of electrostatic potential $\varphi(\vec{r})$, generated by molecular dipoles or quadrupoles. In organic layers sandwiched between  conducting electrodes the potential $\varphi(\vec{r})$ must obey a boundary condition at the electrode surface: here potential should be a constant. Thus, at the electrode there is no electrostatic disorder at all, irrespectively to how disordered is the material in the bulk. This means that the magnitude of the dipolar or quadrupolar disorder increases while going away from the electrode, asymptotically reaching its bulk value.

\begin{figure}[tbph]
\begin{center}
\includegraphics[width=3.2in]{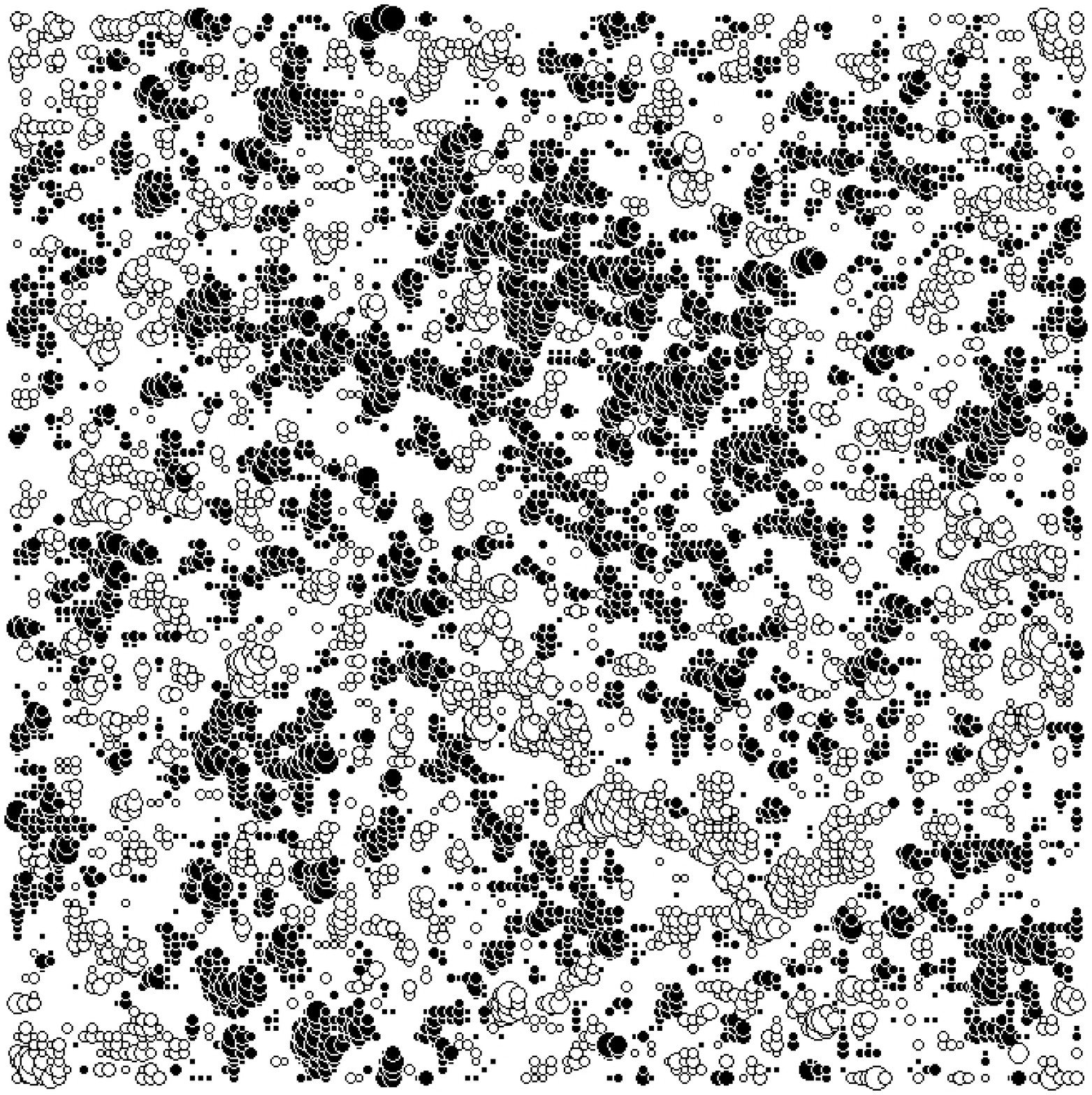}
\end{center}
\centering b
\begin{center}
\includegraphics[width=3.2in]{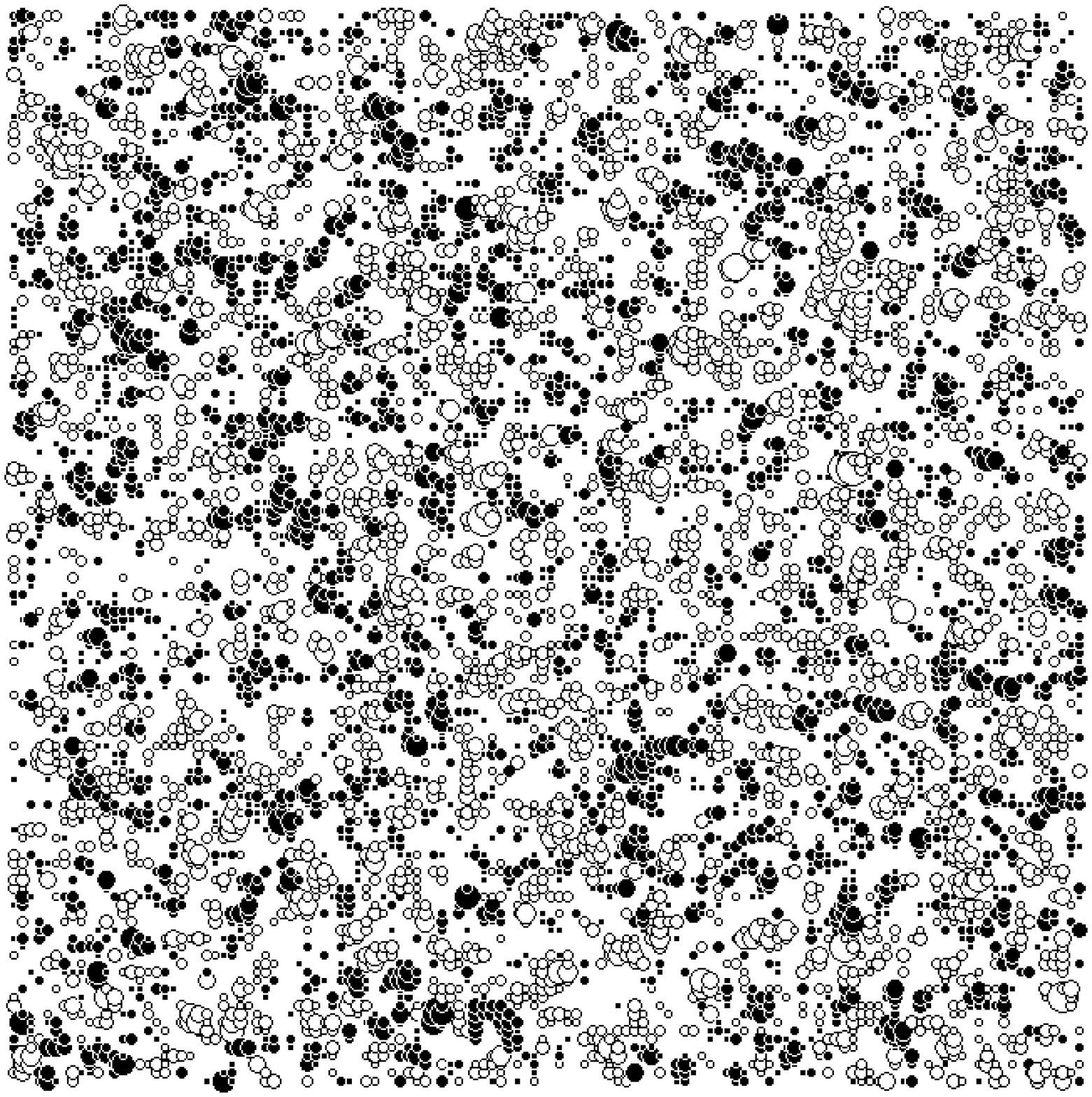}
\end{center}
\centering a
\caption{Distribution of the carrier random energy
$U(\vec{r})$ in the bulk of the organic material (a) and in the organic layer closest to the electrode (b).
All symbols are exactly the same, as used in Fig.\ref{clusters}. Note the significant decrease of the amplitude of the disorder at the electrode and its much less correlated nature.}
\label{electrode}
\end{figure}

Some decrease of the disorder at the surface of organic material is inevitable for any model of organic glass (just because there are more neighbor molecules in the bulk of the material), but the magnitude of the effect in the case of electrostatic disorder is much greater than in the case of short range interactions. For example, for the simple model of the interaction with the nearest neighbors only and simple cubic lattice we have $\sigma^2_{\rm surface}=5/6\sigma^2_{\rm bulk}$, while for the dipolar disorder $\sigma^2_{\rm surface}\approx 0.3\sigma^2_{\rm bulk}$ and rms disorder depends on the distance $z$ from the electrode as \cite{Novikov:033308}
\begin{equation}\label{sigma2(z)_3}
\sigma^2(z)\approx\sigma^2_{\rm bulk}
\left[1-\frac{a_0}{2z}\left(1-e^{-2z/a_0}\right)\right],
\hskip10pt a_0=A_d a
\end{equation}
(here $\sigma_{\rm surface}$ is the magnitude of the disorder in the first layer of organic material, directly adjacent to the electrode). Spatial correlations at the electrode do differ too; this phenomenon has no analogue for the short range disorder. A direct calculation of the correlation function $C(\vec{r})$ near the electrode gives
\begin{equation}\label{z=z'}
    C(z_1,z_2,\vec{\rho})=\sigma^2_{\rm bulk} a_0\left(\frac{1}{r_-}-\frac{1}{r_+}\right),
\end{equation}
where $r^2_\pm=\rho^2+(z_1\pm z_2)^2$ and $\vec{\rho}$ is a 2D vector oriented along the electrode plane \cite{Novikov:949}. Hence, at the electrode the dipolar glass is much less correlated in comparison with the bulk (Fig. \ref{electrode}): $C(z_1,z_2,\vec{\rho})\propto z_1z_2/\rho^3$ for $\rho \gg z_1,z_2$, and clusters are elongated perpendicular to the electrode plane.

Decrease of the disorder at the electrode and change of the spatial behavior of the correlation function have very important implications for the charge injection. In the absence of these phenomena, injection current in organic glasses demonstrates formation of channels where the current density is much greater than the average density \cite{Tutis:161202}. Such channels originate from particular spots at the electrode, where clusters of sites with low energy
facilitates injection. Reduction of the disorder at the electrode and modification of the spatial behavior of the correlation function lead to the more uniform distribution of the injection current over the electrode and dramatically reduce current channeling. This decreases local overheating in a device and improves its performance.

\begin{figure}[tbh]
\begin{center}
\includegraphics[width=2.8in]{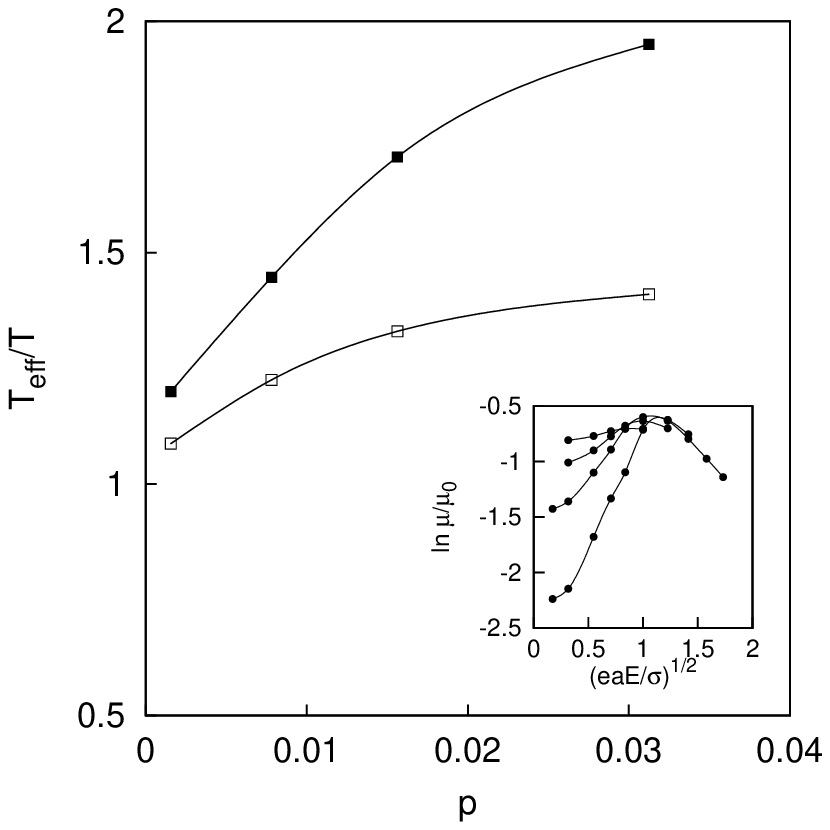}
\end{center}
\centering a
\begin{center}
\includegraphics[width=2.8in]{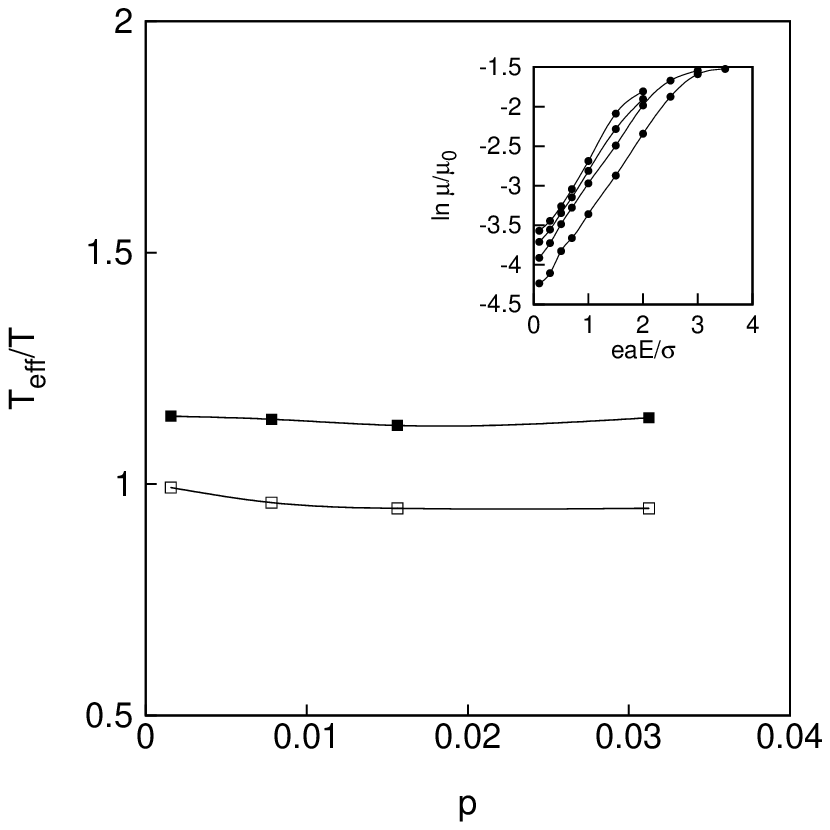}
\end{center}
\centering b
\caption{Dependence of the effective temperature on the carrier density $p$ for $kT/\sigma=0.3$ ($\blacksquare$) and $kT/\sigma=0.4$ ($\square$) for the DG model (a) and GDM (b), correspondingly. Insets show the curves $\mu(E)$ for $kT/\sigma=0.3$ and different $p$:  0.0016, 0.008, 0.016, and 0.032, from the bottom curve upward, correspondingly. For the DG $T_{\rm eff}$ was calculated by fitting the mobility data to Eq. (\ref{eq5}), while for the GDM the mobility data was fitted to Eq. (\ref{mu_GDM}). Lines are provided as guides for an eye.}
\label{Teff}
\end{figure}


In conclusion, we would like to emphasize that by no means the presented calculation gives a full and accurate estimation of the drop of $\sigma$ at the electrode. In fact, as it was mentioned already, there are many other reasons for the variation of magnitude of energetic disorder at the electrode and in many cases the total disorder increases near the electrode. There are three major reasons to consider the particular drop of the electrostatic disorder at the electrode. First, this effect provides a significant contribution to the variation of the magnitude of total disorder, in polar materials it could reach about 50\% of $\sigma_{\textrm{bulk}}$. Second, this effect is universal, it does not depend on particular details of the structure of the material at the electrode (though, of course, an actual magnitude of the drop does depend on the structure of the organic glass near the electrode). And third, it is very unusual, we already noted that for many other contributions we should expect the increase of the disorder at the electrode.

\section{Transport of interacting carriers}
\subsection{Dynamic effects of interaction}
If charge carrier density is not very low, we cannot neglect the Coulomb interaction between carriers. The relevant interaction strength parameter is $U_C=e^2n^{1/3}/\varepsilon\sigma$, where $\varepsilon\simeq 2.5 - 3$ in organic materials and $n$ is a carrier concentration. The maximal value of $U_C$ for $n\simeq 1/a^3$ in organic glasses is $U_C^{\rm max}\simeq 5$. This means that for high concentration of carriers we cannot neglect the inter-charge interaction.

Theoretical studies of the effect of carrier density are not numerous \cite{Yu:721,Pasveer:206601,Zhou:153201}. High density of carriers could affect drift mobility in opposite ways. Small fraction of carriers could occupy deep states, thus providing a possibility for remaining carriers to avoid trapping and acquire much higher mobility. At the same time, charge-charge interactions could provide an additional energetic disorder in the material. This is indeed the case for the simplest model where all charges except one are immovable \cite{Novikov:119,Novikov:191a}. In this case, the greater is the density of static charges (i.e. energetic disorder), the smaller is the mobility.

Usually, in theoretical studies the mean field approximation has been used and charge-charge interaction has been totally neglected \cite{Yu:721,Pasveer:206601}. This means that these studies dealt only with the effect of filling of deep states: it is assumed that the
effects of interaction could be later effectively included via the mobility dependence on the mean local electric field $\left<\vec{E}_{\rm loc}\right>$, which in turn is connected to the mean local charge density $\rho$ by the Poisson equation
\begin{equation}\label{Poisson}
{\rm{div}} \left<\vec{E}_{\rm loc}\right> =
\frac{4\pi}{\varepsilon} \rho.
\end{equation}
This line of reasoning totally neglects dynamic correlations. In addition, quite frequently true quasi-equilibrium mobility is formed by the averaging over large domains of the disordered material (see, for example, Ref. \cite{Pasveer:206601}). In this situation the very conception of a local (but uniform in space)
mobility  is invalid and detailed simulation of the transport of interacting carriers is unavoidable.

Effect of the interaction for the hopping transport in the DG model and GDM has been studied in recent paper \cite{Novikov:740}. It was found that the spatial correlation manifests itself even in the case of high carrier density. Indeed, a general tendency for the DG model is that transformation of the mobility curve with the increase of average fraction of the occupied sites $p=na^3$ resembles the corresponding transformation of the curve with the increase of $T$ (compare insets in Fig. \ref{Teff}, for example, with Fig. 1 in Ref. \cite{Novikov:4472}): with the increase of $p$ mobility becomes greater and the slope of $\mu(E)$ curve becomes smaller. This is not the case for the GDM: here only mobility curve moves upward but the slope remains approximately constant.

This difference could be easily understood. It was noted that the field dependence of $\mu$ in the GDM is governed by the carrier escape from deep states to the nearest sites having much higher energy, and the field-induced shift of site energies leads to \eq{mu_GDM_simple}. Random charge distribution provides a smooth random energy landscape superimposed on the intrinsic disorder, but typical additional variation of energy at the scale $a$ is negligible for small $p$. Hence, estimation (\ref{mu_GDM_simple}) remains valid and the slope of the mobility curve does not depend on $p$.

Situation in the DG model is different: here mobility field dependence is governed by the carrier escape from critical clusters, as described in Section \ref{sect_transport}. If we increase the density of
carriers, then at first they fill these critical traps, because the release time is maximal here. Hence, transport of more mobile carriers is governed by clusters with the size that differs from $r_{\rm cr}$ (it is smaller). This means that the effective critical size $r^{\rm eff}_{\rm cr}$ depends on $p$. According to Eq. (\ref{critical}), this is equivalent to the introduction of the effective temperature $T_{\rm eff}$, depending on $p$, and $T_{\rm eff}$ grows with $p$. This conclusion is in good agreement with Fig. \ref{Teff}: while for the GDM $T_{\rm eff}$ does not depend on $p$ and is very close to $T$, for the DG model $T_{\rm eff}$ monotonously grows with $p$.

\begin{figure}[floatfix]
\begin{center}
\includegraphics[width=3.1in]{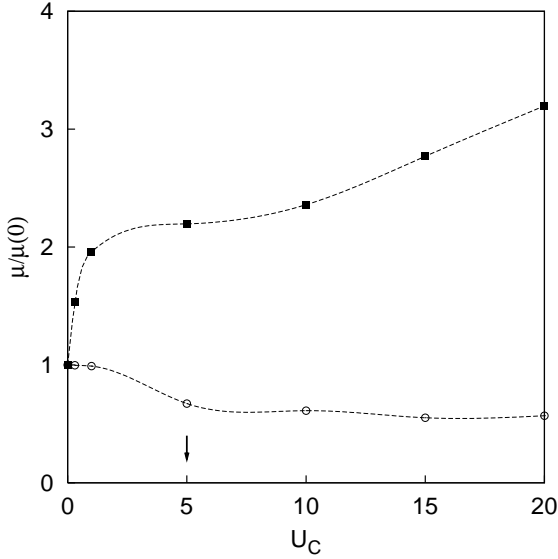}
\end{center}
\caption{ Dependence of the carrier mobility on
the effective strength of the charge-charge interaction
$U_C=e^2/\varepsilon a \sigma$ for $p=0.008$,
$kT/\sigma=0.3$, and $eaE/\sigma=0.1$. Black squares show the data for the DG model and empty circles show results for the GDM. For a typical disordered organic material $U_C \simeq 5$. Here $\mu(0)$ is the mobility for $U_C=0$.} \label{Int_effect}
\end{figure}

It was found also that for the DG model in the case of moderate $p \leq 0.1$ carrier drift mobility increases with $p$, exactly as in the case of non-interacting carriers \cite{Yu:721,Pasveer:206601}. Yet modification provided by the interaction is still significant
(see Fig. \ref{Int_effect}). There is a striking difference between the effects of interaction (i.e., carrier-carrier repulsion) on the mobility in the DG model and GDM. In the DG model repulsion between carriers makes mobility even greater than in the case of no interaction, while for the GDM the opposite situation takes place. This difference agrees with the cluster structure of the DG: if a carrier is trapped by some  valley of the energetic landscape, then the whole valley with many sites having low energies becomes blocked for other carriers because of repulsion. Thus, filling of the deep states is much more effective in correlated landscape if carrier repulsion is taken into account. This is the reason for the increase of carrier mobility in DG, in comparison with the case of non-interacting carriers. No such effect takes place for the GDM, and here, evidently, the effect of charge-induced energetic disorder is responsible for the decrease of mobility in comparison to non-interacting case.

\begin{figure}[floatfix]
\begin{center}
\includegraphics[width=3.4in]{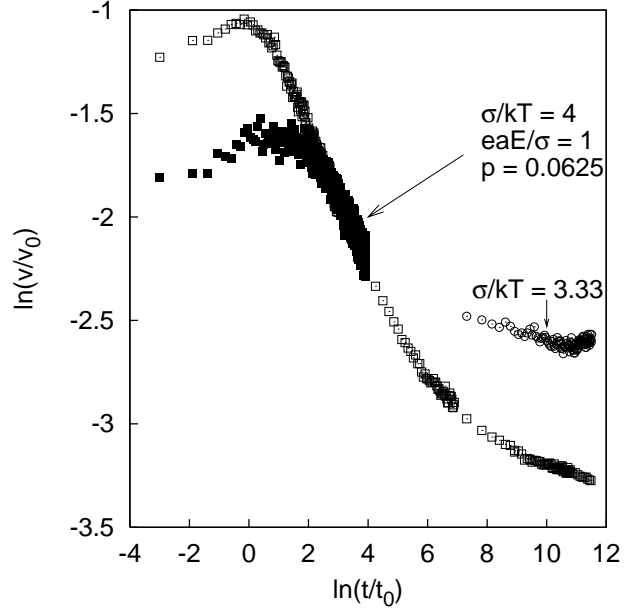}
\end{center}
\caption{ Relaxation of the mean carrier velocity in the GDM for $\sigma/kT=4$, $eaE/\sigma=1$, and $p=0.0625$ (squares). Empty squares show relaxation for random initial locations, while filled squares show relaxation for the case, when initial locations have been taken at the minimal energy positions. Circles shoe the
relaxation for higher temperature $\sigma/kT=3.33$. Time and velocity scales are $t_0=1/\nu_0$ and $v_0=a/t_0$, correspondingly.}
\label{relaxation}
\end{figure}

This particular result disagrees with the result of a recent paper by Zhou \textit{et al} \cite{Zhou:153201}. They found that in the GDM carrier interaction \textit{enhances} mobility in comparison to the case of no interaction if $\sigma/kT \gg 1$. This is opposite
to our findings. Quite probably, the disagreement stems from the under-relaxation of the initial (random) carrier configuration used in Ref. \cite{Zhou:153201}; the relaxation process is pretty slow for interacting carriers if $\sigma/kT \gg 1$. We cannot make more
detailed comparison because typical relaxation times are not provided in Ref. \cite{Zhou:153201} (in fact, even the strength of carrier repulsion is not provided). Our data indicates, for example, that for $\sigma/kT=4$ relaxation is not completely over even for $t/t_0=1\times 10^5$ (see Fig. \ref{relaxation}); at that time carrier has already traveled in the field direction the distance of $\simeq 4\times 10^3 a$.

Remarkable feature of Fig. \ref{relaxation} is a universality of the late relaxation stage. Very early relaxation is different for the initial random distribution and minimal energy distribution
(where every carrier was placed at the site where the total energy, provided by the intrinsic disorder and all previously added carriers, has a minimum), but after $t/t_0\simeq 10$ relaxation curves merge into a single curve.

\subsection{Effects of interaction: comparison with experiment}

One can suggest that carrier transport in organic field-effect transistors (OFETs) should be a natural choice for comparison of the simulation results with experiment \cite{Horowitz:1946,Sirringhaus:2411}. Estimation of the carrier density in OFETs show that the density as high as $3\times 10^{19}$ cm$^{-3}$ could be achieved \cite{Tanase:216601}, that for $a\approx 1$ nm corresponds to $p\approx 0.03$. Experimental data for the particular OFET should be compared with the TOF data for a sandwich device having transport layer of the same material; in this way we could use transport characteristics (e.g., $\sigma$), relevant for the intrinsic disorder in the material. Quite frequently, OFETs demonstrate mobilities much higher that the mobilities measured in TOF experiments, and usually mobility increases with the increase of $p$ \cite{Tanase:216601}. This fact is in general agreement with the model studied in this paper.

However, careful analysis reveals much more complicated situation. Indeed, in many aspects OFETs are very far away from the model, considered in the current study. First of all, in OFETs transport occurs in a thin layer, close to the gate insulator. Quite probably, especially in polymer devices, structure of this layer differs from the structure of the same material in the bulk (polymer chains could be arranged in a special way at the gate
insulator surface). This arrangement could provide more ordered structure with less degree of energetic disorder, thus mobility should be enhanced near the interface, but accumulation of surface defects and impurities at the interface could lead to the decrease of the mobility. Next, there is a clear indication that the roughness of the organic semiconductor/dielectric interface affects carrier mobility \cite{Chua:1609}. At last, the very nature of a gate dielectric (specifically, its polarity) affects carrier mobility in OFETs, because a random orientation of polar groups in the vicinity of a transport layer induces an additional energetic disorder in semiconductor \cite{Veres:199,Veres:4543}. In short, there are a lot of reasons to believe that transport properties of OFETs are too complicated to be directly compared with the results of this study. We can only state that a significant increase of the carrier mobility with the increase of carrier density in carefully manufactured OFETs does not contradict the results of our study.

\section{Conclusion: future development}

We may conclude that the spatially correlated electrostatic energetic disorder is a dominant factor directly dictating major features of the hopping charge transport and injection in amorphous organic materials. Strong spatial correlation is an inevitable property of the random energy landscape, created by long range electrostatic sources. It is worth to note that this inevitability is still not recognized by a significant part of the community. Sometimes one can read in papers that "our analysis does not indicate that there is a need to assume a certain spatial energy correlation" (see, e.g. the paper \cite{Pasveer:206601}). Typically, it is believed that the non-correlated distribution of random energies is something natural (probably, because of the simplicity of the conception), but the correlated distribution needs the specific reasons to occur. We hope that the discussion in this paper demonstrates that in organic glasses quite the opposite is true: they are correlated media by very nature. If, for some reason, there is a need to suggest a non-correlated energetic disorder for charge carriers, then the very existence of such disorder in organic glasses is very difficult (probably, impossible) to justify.

At the same time, our knowledge of some important aspects of the hopping transport in correlated landscape is not sufficient. We may mention the almost absolute absence of reliable (all the more so, exact) analytical results for the 3D transport, scarcity of theoretical results on the shape of the current transients, and other open problems.

We would like especially emphasize the ultimate deficit of the experimental data on the local orientation order in organic glasses: the local order could change the correlation function for short distances and, hence, change the mobility field dependence for strong fields. Our current knowledge in this area can be estimated using recent papers  \cite{Qi:9455,Senker:7592,Reichert:542,Bartolotta:4798},
it is absolutely insufficient for the reliable consideration of transport problems. Study of the transport of interacting carriers is still in its infancy. Reliable experimental evidence for the modification of electrostatic disorder at the electrode was not provided.

Several recent papers advanced a program of so-called multiscale modeling of charge transport in amorphous organic materials \cite{Kirkpatrick:227402,Nelson:1768,Ruhle:134103}. According to this program, the very structure of the organic glass is simulated, using some variant of the molecular dynamics or related approach, then the relevant parameters (positions of energy levels, hopping probabilities) are calculated for every particular  particular realization of the structure of the glass, and then the Monte Carlo simulation of the hopping transport is carried out. Using this approach, it was possible in some cases to calculate the mobility value, pretty close to measured in experiments (the difference is about one order of magnitude) \cite{Nelson:1768}. At the same time, typical calculated mobility field dependences are much less steep, than the experimental ones. Probably, the reason for this difference is insufficient size of the basic simulation sample (about $10^3$ molecules). Obviously, limitations of the multiscale simulation are mostly determined by the achieved computer performance which is constantly increasing. Hence, this approach seems to be a very promising line of future investigation. In addition, it could provide a valuable information on the local structure of organic glasses.

This short and, by no means, exhausting list of open problems and possible directions of further investigation could serve as a proof for the reader that the study of hopping charge transport in organic materials will be a vibrant area of research for the observable future.

\begin{acknowledgments}
I am grateful for valuable discussions to A.V. Vannikov. A.P. Tyutnev, and L.B. Schein. Partial financial support from the RFBR grants 10-03-92005-NNS-a and 11-03-00260-a  is acknowledged.
\end{acknowledgments}

\bibliographystyle{apsrev4-1}
\bibliography{D:/Projects/TeX/BibTeX_Primer/full_db2short}

\end{document}